\documentclass[onecolumn,superscriptaddress,aps,preprintnumbers,amsmath,amssymb]{revtex4-1}
\usepackage{graphicx}
\usepackage{dcolumn}
\usepackage{bm}
\usepackage{hyperref}
\usepackage{algorithm2e}
\usepackage{xfrac}
\usepackage{braket}
\usepackage{mathtools,amssymb,lipsum} 
\usepackage[utf8]{inputenc}
\usepackage{color,soul}
\begin{document}
\title{Manuscript Title:\\with Forced Linebreak}
\author{Chumki Nayak}
\thanks{Present Address: Center for van der Waals Quantum Solids, Institute for Basic Science (IBS), Pohang, Korea}
\affiliation{Department of Physical Sciences, Bose Institute, 93/1, Acharya Prafulla Chandra Road, Kolkata 700 009, India}
\author{Suvadip Masanta}
\affiliation{Department of Physical Sciences, Bose Institute, 93/1, Acharya Prafulla Chandra Road, Kolkata 700 009, India}
\author{Shubhadip Moulick}
\thanks{These two authors contributed equally.}
\affiliation{Department of Condensed Matter and Materials Physics, S.N. Bose National Centre for Basic Sciences, Kolkata 700106, India}
\author{Manotosh Pramanik}
\thanks{These two authors contributed equally.}
\affiliation{Department of Physical Science, Indian Institute of Science Education and Research Kolkata, Mohanpur, Nadia, 741246, West Bengal, India}
\author{Atanu Kabiraj}
\affiliation{Department of Physics, School of Basic Sciences, Indian Institute of Technology Bhubaneswar, Jatni-752050, Khordha, Odisha, India}
\author{Satchidananda Rath}
\affiliation{Department of Physics, School of Basic Sciences, Indian Institute of Technology Bhubaneswar, Jatni-752050, Khordha, Odisha, India}
\author{Sukanya Ghosh}
\affiliation{Department of Physics, Central University of Kashmir, Jammu and Kashmir, India}
\author{Atindra Nath Pal}
\affiliation{Department of Condensed Matter and Materials Physics, S.N. Bose National Centre for Basic Sciences, Kolkata 700106, India}
\author{Bipul Pal}
\affiliation{Department of Physical Science, Indian Institute of Science Education and Research Kolkata, Mohanpur, Nadia, 741246, West Bengal, India}
\author{Achintya Singha}
\email{achintya@jcbose.ac.in}
\affiliation{Department of Physical Sciences, Bose Institute, 93/1, Acharya Prafulla Chandra Road, Kolkata 700 009, India}
\begin{abstract}
Alloying offers an effective way to improve the functionality of transition metal dichalcogenides (TMDCs) in both fundamental research and optoelectronic applications, as it allows for engineering their electronic and optical properties. This study investigates the optoelectronic properties of CVD-synthesized alloy MoSSe, which exhibits an inherent out-of-plane dipole moment, arising from asymmetry in S and Se atoms on either side of the Mo layer, as confirmed by piezoelectric force microscopy, polarization-resolved second harmonic generation studies and theoretical first-principles calculations. Time-resolved photoluminescence measurements reveal an extended exciton radiative recombination lifetime in MoSSe, attributed to electron-hole wavefunction separation by the dipole moment, which improves photodetection by facilitating enhanced electron-hole separation before recombination. The device demonstrates significant responsivity over broad spectral range. By employing the photogating effect, the device response can be switched from slow to fast modes. These findings are further supported by illumination intensity-dependent photoluminescence and Raman measurements, underscoring the potential of polar TMDCs in future optoelectronic devices.
\end{abstract}
\title{Intrinsic Electric Field Driven High Sensitive Photodetection in Alloy TMDC MoSSe}
\maketitle

\section{Introduction}
Photodetectors are indispensable devices that convert incident light into electrical signals and used in diverse applications, ranging from telecommunications to biomedical imaging \cite{Konstantatos2018, Koppens2014, Zhang2013}. 
The discovery of two-dimensional (2D) materials provide a fruitful platform for scaling down the device dimensions and achieving multifunctional photoresponsive devices 
\cite{doi:10.1021/acsnanoscienceau.2c00017, Roy2013}. Transition metal dichalcogenides (TMDCs), a prominent group among 2D materials, are attracting considerable attention for developing efficient photodetectors due to their layer-dependent electronic and optical properties \cite{doi:10.1021/nl903868w, Schaibley, Xiaodong, Chen2023, Radisavljevic2011}. 
However, challenges like short exciton lifetimes, strong exciton-exciton interactions at room temperature, and a narrow spectral window hinder the effectiveness of TMDCs in nanophotonic devices
\cite{doi:10.1021/acsami.1c24308, Dhyani2017}. Consequently, various approaches including bandgap engineering, defect modification, and heterostructure formation have been employed to enhance light interaction with TMDCs \cite{Masanta, Liu, Jiang, Liang, Novoselov, PhysRevB.92.155403}.
\par
Alloy engineering offers a powerful approach to tailor the band gap and optimize the electronic and optical properties of 2D TMDCs \cite{Nayak, Bera, doi:10.1021/acsnano.5b02506, https://doi.org/10.1002/adma.201901405, Mukherjee2022}. The formation of alloy TMDC, denoted as  MX$_{2x}$Y$_{2(1-x)}$, involves substituting some X chalcogen atoms in the parent TMDC (MX$_2$) with Y chalcogen atoms, enabling a tunable and broad emission spectrum \cite{C5NR02515J}. By lowering the density of localized defect states and converting deep-level defect to shallow-level ones, alloys reduce non-radiative recombination rates \cite{D0NA00202J, https://doi.org/10.1002/adma.201901405}. Beyond defect mitigation, alloy TMDCs exhibit an additional unique advantage stemming from their non-centrosymmetric crystal structure. The spatial arrangement of chalcogen atoms with differing electronegativities on opposite sides of the transition metal atoms induces an out-of-plane electric field. This intrinsic field has been predicted theoretically to influence the overlap of out-of-plane wavefunctions for electrons and holes, thereby modulating excitonic properties \cite{PhysRevB.99.115316, C8NR04568B}. These findings have expanded opportunities for designing highly sensitive photodetector devices based on alloy materials. In the light of the recent reports, there are ample opportunity to explore the optoelectronic properties of alloy TMDC, particularly concerning synthesis method, spectral range, and device speed \cite{D0NA00202J, https://doi.org/10.1002/adma.201901405}. Moreover, a deeper understanding of the underlying mechanism could enable the design of photodetectors tailored to specific applications.
\begin{figure*}[b]
\centering
\includegraphics[width=1\linewidth]{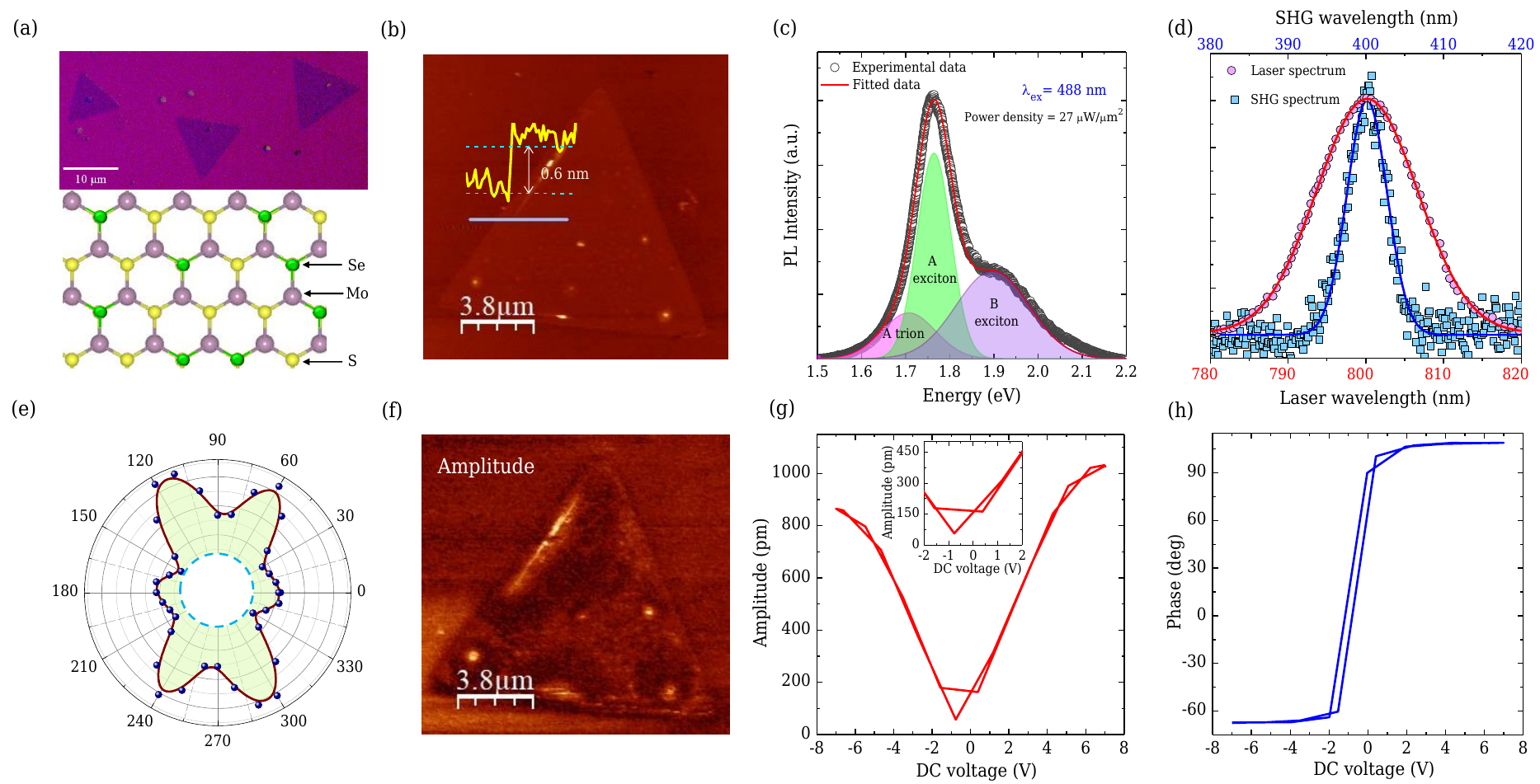}
\caption{Primary characterizations of as-synthesized MoSSe sample. (a) Optical image (top) and top view of the atomic structure (bottom), (b) AFM topography, (c) PL spectrum of MoSSe. (d) Normalized SHG along with corresponding input laser spectra. (e) Polar plots of the polarization-resolved SHG intensity, measured with the detection polarizer maintained parallel to the input laser polarization (0 degree corresponds to S polarized input). (f) PFM amplitude image, (g) and (h) PFM amplitude and phase as a function of DC bias voltage of the monolayer MoSSe, respectively. Inset of Figure (g) shows the appearance of butterfly-shaped feature within the DC voltage range of -0.5 V to +1 V}.
\label{FIG. 1.}
\end{figure*}

In this study, we successfully synthesized the alloy TMDC MoS$_{2x}$Se$_{2(1-x)}$ ($x =$ 0.6) using a chemical vapor deposition (CVD) method. Piezoelectric force microscopy (PFM) and polarization-
resolved second harmonic generation (SHG) measurements along with the first-principles calculations based on density functional theory (DFT) confirm the presence of out-of-plane electric field in the monolayer sample. The electric field controls the dynamics of electron-hole recombination in the TMDC alloy material. Additionally, time-resolved photoluminescence (TRPL) measurements revealed an extended exciton lifetime, which enhances the probability of charge separation and improves overall carrier collection. Our synthesized MoSSe based device exhibits significant photosensing behavior across a wide spectral range (400-1100 nm), with increased responsivity and detectivity at low power density. We explored the tunability of these properties under optical and electrical control. The response speed was tailored through the photogating effect, showing potential for applications ranging from fast photodetectors to optoelectronic memory devices.
The combined effects of the inherent electric field and photogating contribute to enhanced photosensing behavior, supporting the development of highly sensitive, broadband photodetectors. Furthermore, the modulation of PL and Raman properties with varying illumination intensity provides insight into the fundamental principles driving the material's optoelectronic properties.

\begin{figure*}[b]
\centering
\includegraphics[width=0.6\linewidth]{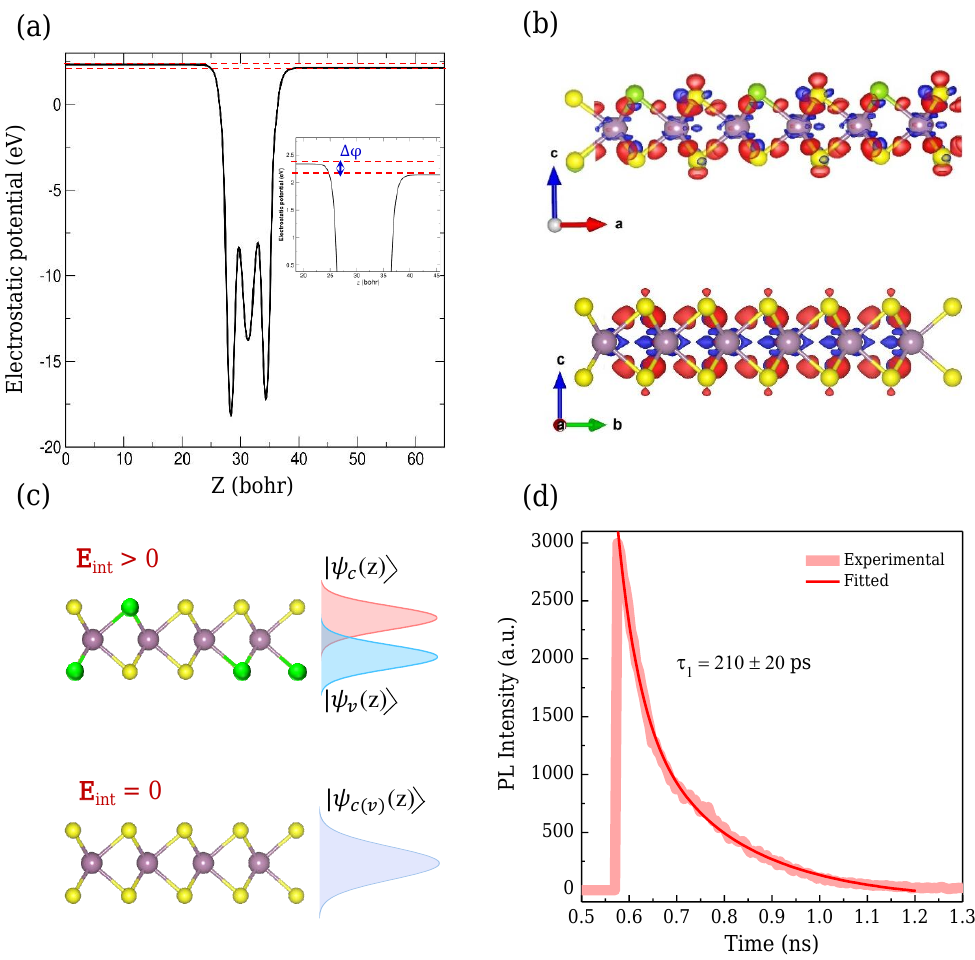}
\caption{Effect of out-of-plane electric field in TMDCs. Electrostatic potential with respect to the vacuum level is plotted along the direction perpendicular to the plane of (a) monolayer MoSSe, 
 (inset shows the zoom view of the vacuum potential between the two terminal chalcogen layers). The red dashed lines show the vacuum potential between the two terminal chalcogen layers. (b) Partial charge density difference for  monolayer MoSSe (top) and MoS$_2$ (bottom), respectively. The red and blue isosurfaces show electron accumulation and depletion, respectively, drawn at =  0.01 e/bohr$^3$. The purple, yellow and green atoms show Mo, S and Se, respectively. (c) Schematic representation of out-of-plane wavefunctions for MoSSe (top) and MoS$_2$ (bottom). The presence of built-in electric field ($E_{int}$) in MoSSe significantly reduces the overlap and separates the center of the out-of-plane wavefunctions of electrons and holes and (d) TRPL spectrum of A exciton of the MoSSe.}
\label{FIG. 4.}
\end{figure*}

\section{Results and Discussion}
MoSSe crystals,  synthesized by the CVD method, are detailed in the supplemental material (Section I). Triangular samples are randomly distributed on the SiO$_2$/Si substrate, as shown in the optical image (Figure 1a, top). The top view of the MoSSe atomic structure is schematically presented in Figure 1a (bottom). The sample selected for device fabrication was first analyzed by atomic force microscopy (AFM) imaging, and the extracted height profile from topography confirms it as a monolayer (Figure 1b and its inset). Compositional analysis of Mo, S, and Se was performed using energy-dispersive X-ray spectroscopy (EDX). The obtained EDX spectrum and the calculated atomic percentages are shown in Figure S2a. The PL spectrum is presented in Figure 1c. Strong Coulomb interactions lead the electron in the conduction band and the hole in the valence band to bind, forming an exciton in the direct band-to-band transition.
Two distinct excitonic A and B bands appear in the PL spectrum, originating from the splitting of the valence band maximum ($\sim$ 140 meV) and conduction band minimum ($\sim$ 10 meV) due to strong spin-orbit coupling involving Mo-d electrons \cite{Shan}. In the lower energy region of the PL spectrum, an additional peak appears due to A trion emission. The energies associated with the A trion, A, and B excitons are 1.74 eV, 1.77 eV, and 1.91 eV, respectively.
Figure S2b displays the Raman spectrum of the sample, with prominent Raman modes A$^1_1$, E$^3_2$, E$^2_1$, and A$^2_1$ observed at 272 cm$^{-1}$, 368 cm$^{-1}$, 384 cm$^{-1}$, and 405 cm$^{-1}$, respectively, in agreement with previous findings \cite{PhysRevB.109.125306, PhysRevB.109.115304}.
\par
To study nonlinear optical response, we performed SHG measurements on monolayer MoSSe (for more details, see supplementary, Section III). For an input laser spectrum centered at 800 nm, the detected SHG spectrum was centered at 400 nm (Figure 1d), as expected. The quadratic dependence of the SHG intensity on the input laser intensity was confirmed in Figure S4a. The polar plots of the polarization-resolved SHG intensity (Figures 1e and S4b) display a six-lobed pattern, obtained by rotating the input laser polarization using a half-wave plate, while maintaining the detection polarizer parallel or perpendicular to the input polarization \cite{doi:10.1021/acsnano.4c02854, https://doi.org/10.1002/adma.201701486}. This pattern reflects the broken inversion symmetry and the presence of an in-plane dipole moment. Additionally, the non-zero, isotropic background SHG intensity, marked by the blue dotted circle in the aforementioned figures, indicates the presence of an out-of-plane dipole moment in the monolayer MoSSe \cite{doi:10.1021/acsnano.4c02854, https://doi.org/10.1002/adom.202300958}. The low SHG intensity near and at 0 $^\circ$/180 $^\circ$ angle arises probably due to unintentional presence of strain in the 2D layered materials \cite{Mennel2018, 10.1063/1.5051965, doi:10.1021/acs.nanolett.0c00694}.

\par
We conducted vertical mode PFM measurements, to further experimentally confirm the presence of an out-of-plane dipole moment resulting from the structural asymmetry of monolayer MoSSe. In our previous study, we detailed the PFM measurement techniques for the MoSSe alloy samples \cite{PhysRevB.109.115304}. The piezoresponse observed in the PFM measurements is based on the inverse piezoelectric effect, where applying a voltage induces a mechanical strain in the material. The sample under the cantilever tip exhibits surface deformations, expanding or contracting based on whether its local polarization (linked to an intrinsic vertical dipole moment) aligns parallel or antiparallel to the applied AC field. Figures 1f, and S5b depict the piezoelectric amplitude domain, and phase images, respectively. The piezoelectric contrast between the sample (triangular region) and the substrate in Fig. 1f, along with the piezoresponse amplitude profile shown in Fig. S5a along the marked line in Fig. 1f, clearly demonstrates the piezoelectric response originating from the sample. In this measurement, the polarization orientation (up or down) can be modified by applying a DC field. This is visualized through the plots of amplitude versus DC voltage and phase versus DC voltage in the presence of an AC probing voltage ($V_{AC}$). In the amplitude plot (Figure 1g), the characteristic butterfly-shaped feature (inset) appears within a small DC voltage range. The phase plot shows clear evidence of hysteresis, with sharp changes of approximately 167$^\circ$ in polarization direction (Figure 1h). The deviation from the ideal 180$^\circ$ phase reversal can be attributed to factors such as sample imperfections and experimental conditions. The twin features of hysteresis in both the amplitude and phase plots, along with the evidence of local polarization switching, confirm the  presence of an intrinsic out-of-plane electric dipole moment in the sample \cite{Lu2017, Xie, Xue, Rao}. The piezoelectric coefficient, $d_{33}$, is calculated using the formula $A_{1w} = d_{33}V_{AC}Q$, where $A_{1w}$ represents extracted piezoresponse amplitude. For the cantilever quality factor ($Q$) of 100, $d_{33}$ is determined to be 1.58 $\pm$ 0.03 pm/V. It is important to note that these $d_{33}$ values are qualitative, as the effective piezoelectricity of the samples may be influenced by small variations in their electrical properties. Furthermore, PFM measurements on the SiO$_2$/Si substrate (Fig. S6) reveal no characteristic piezoresponse (such as polarization switching), indicating that the piezoelectric response arises from the monolayer sample \cite{PhysRevB.109.115304}.

\begin{figure*}[t]
\centering
\includegraphics[width=0.95\linewidth]{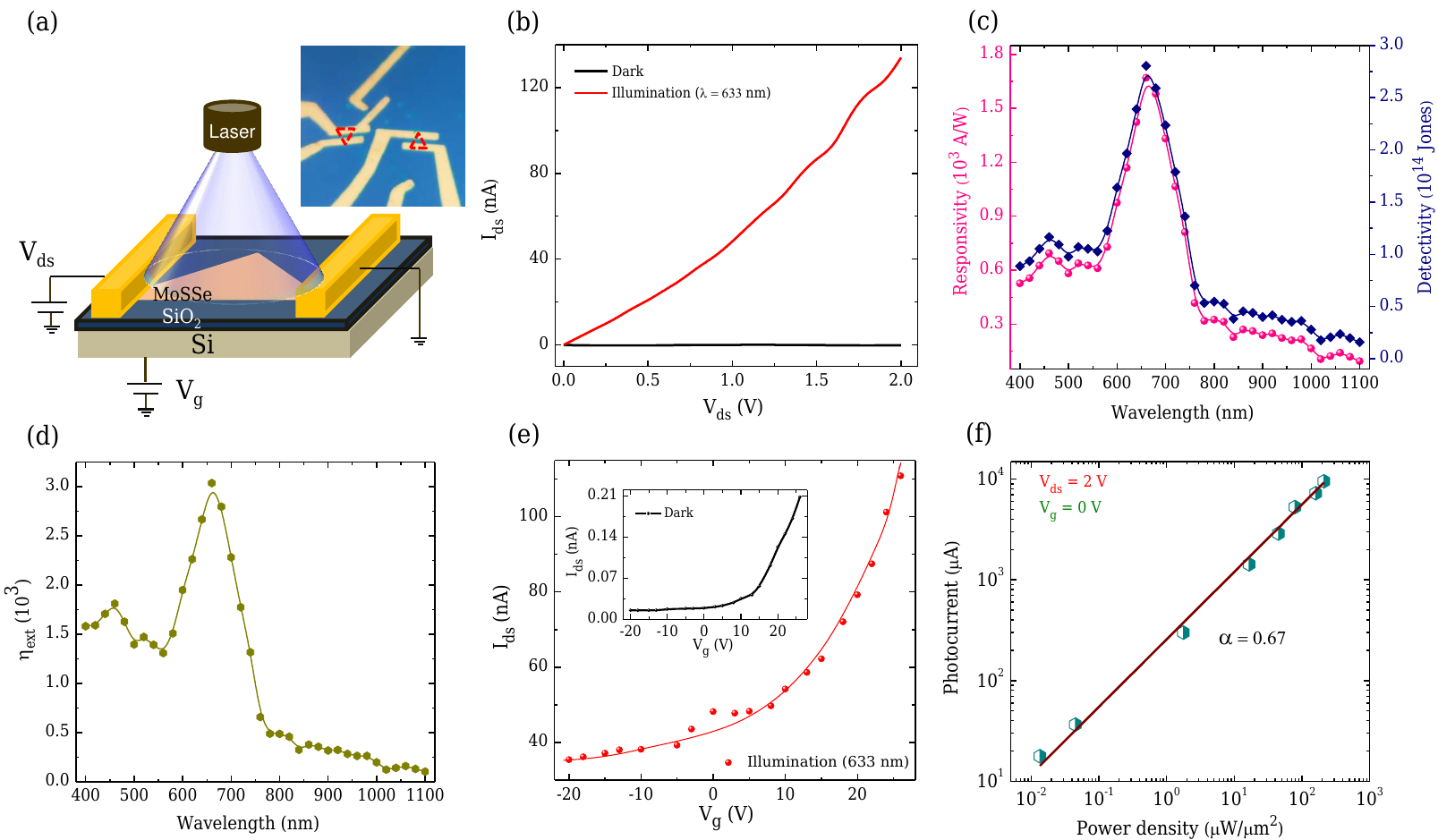}
\caption{Photosensing behavior of MoSSe photodetector.(a) Schematic representation of MoSSe FET device structure. Inset shows optical image of as synthesized device. (b) I-V characteristics of the device under dark (black line) an illumination (633 nm) conditions (red line) at V$_{ds} =$ 0 V to 2 V. (c) Responsivity (pink curve), detectivity (blue curve) and (d) external quantum efficiency ($\eta_{ext}$) at V$_{ds} =$ 2 V over the broad spectral region (400-1100 nm). (e) Transfer characteristics of the device under dark (inset) and illumination conditions with V$_{ds} = $1 V and (f) logarithmic representation of photocurrent as a function of power density.}
\label{FIG. 4.}
\end{figure*}
\par

We have also performed first-principles calculations to gain a deeper understanding of the electronic structure and built-in polarity of the samples. The structural asymmetry in the MoSSe monolayer induces a dipole moment and generates a built-in out-of-plane electric field (E$_{int}$), leading to a vacuum potential energy shift between the two sides of the structure, as shown in Figure 2a (see inset also). In contrast, the absence of such a potential energy shift in monolayer MoS$_2$ confirms the lack of an intrinsic electric field (Figure S7).
The presence of this E$_{int}$ in MoSSe is further supported by the appearance of a Rashba spin–orbit interaction near the point $\Gamma$, which is absent in monolayer MoS$_2$. The fully relativistic band structures of monolayer MoSSe and MoS$_2$, including spin–orbit coupling, along the high-symmetry directions of the Brillouin zone and their enlarged views near the $\Gamma$ point, are shown in Figure S8. To further illustrate the impact of structural asymmetry, the partial charge density difference plots for both materials are presented in Figure 2b. The red and blue lobes show electron accumulation and depletion, respectively, in the monolayer. For, monolayer MoSSe, the electronegativity difference between S (2.58) and Se (2.55) causes asymmetric charge redistribution,  increasing electron accumulation in the vicinity of more electronegative S than Se. In contrast, monolayer MoS$_2$ exhibits a symmetric charge distribution due to the identical chalcogen atoms on both sides. The out-of-plane E$_{int}$ in MoSSe separates the electron and hole along the vertical direction (Figure 2c, top), their overlap decreases, and thereby weakening their interaction. This reduction in overlap ultimately leads to a lower exciton binding energy \cite{10.1063/5.0095203, PhysRevB.98.165424}.
\begin{figure*}[t]
\centering
\includegraphics[width=0.8\linewidth]{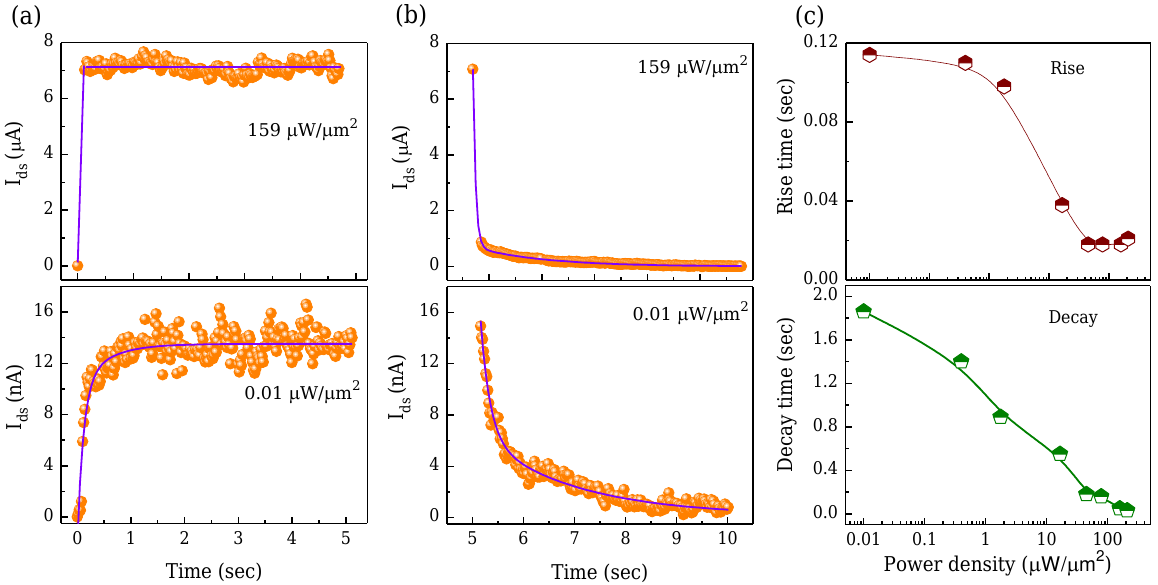}
\caption{Temporal response of MoSSe photodetector. (a) Rise (b) decay response at two illumination intensities (159 and 0.01 $\mu W/\mu m^2$) and (c) semi-logarithmic representation of extracted rise and decay times as a function of illumination intensity.}
\label{FIG. 4.}
\end{figure*}
In contrast, a system without an internal electric field, such as pristine MoS$_2$,  exhibits a higher overlap of the electron and hole wavefunction (Figure 2c, bottom). 
Consequently, MoS$_2$-like materials promote lower electron-hole pair separation prior to recombination, unlike alloy TMDC materials. 
According to Fermi-Golden rule, the exciton lifetime  ($\tau$) can be expressed  as \cite{C8NR04568B, PhysRevB.93.045407},
\begin{equation}
\tau^{-1}= \frac{2\pi}{\hbar} {\vert <b\vert \hat{H} \vert a> \vert}^{2} \delta (E_b-E_a)
\end{equation}
where, $\hat{H}= \frac{e}{mc} \boldsymbol{L}.\boldsymbol{P}$ represents the interaction between the light field's vector potential 
$\boldsymbol{L}$ and the electron's momentum $\boldsymbol{P}$. Here, e denotes the charge of the electron and c is the speed of light. $\vert a>$ and $\vert b>$ correspond to the initial exciton state and the final single-photon state of the system, having energies $E_a$ and $E_b$, respectively. Considering a small exciton distribution around the Dirac point and the momentum operator $\boldsymbol{P}$ acting on the Bloch component of the exciton wavefunctions, the exciton radiative lifetime can be written as \cite{C8NR04568B},
\begin{equation}
\tau \propto \vert <\psi_{c}(z)\vert \psi_{v}(z)> \vert ^{-2}
\end{equation}
where, $\psi_{c(v)}(z)$ represents the out-of-plane wavefunction of the electron (hole).
This expression indicates that a reduced overlap of out-of-plane wavefunctions between the electron and hole extends the exciton radiative lifetime in MoSSe. To explore this effect, we performed TRPL spectroscopy of monolayer MoSSe, as depicted in Figure 2d. Our measurements reveal an A exciton lifetime of 210 $\pm$ 20 ps, in contrast to the exciton lifetime of 43 $\pm$ 5 ps reported for MoS$_2$ \cite{doi:10.1021/acs.nanolett.0c03412}. Another parent TMDC,  MoSe$_2$, exhibits an even shorter exciton radiative lifetime ($\leq$ 3 ps) \cite{PhysRevB.93.205423, Wang2015}. Notably, T. Zheng et al. reported an exciton lifetimes of 134 $\pm$ 10 ps for Janus  MoSSe \cite{doi:10.1021/acs.nanolett.0c03412}. Thus, the extended exciton lifetime observed in the alloy MoSSe represents an improvement over its parent TMDCs and is comparable to Janus materials. This property could enhance the potential for improved photosensing behavior in the alloy materials, motivating further investigation of the optoelectronic properties of MoSSe-based devices \cite{C8NR04568B}.

Figure 3a presents a schematic diagram of the MoSSe based FET device, with an optical image of the actual device (inset). The I-V characteristics of the device (V$_{ds} =$ 0 V to 2 V) are presented in Figure 3b under dark and illumination conditions ($\lambda$ = 633 nm) at a power density of 0.2 $\mu W/\mu m^2$. The device's large illumination current relative to the dark highlights its potential for photodetection applications. We investigated spectral response of the device over a broad wavelength range (400 nm-1100 nm) at low power density (191 fW/$\mu m^2$) with V$_{ds}$ set at 2 V. To evaluate the photodetector's performance, we calculated the  photoresponsivity (R$_\lambda$), detectivity (D$_\lambda$) and external quantum efficiency ($\eta_{ext}$). The R$_\lambda$ using the formula \cite{Nayak, https://doi.org/10.1002/smll.201802593},
\begin{equation}
 R_\lambda = \frac{I_{ph}}{{P_\lambda}{A}}   
\end{equation}
where, I$_{ph}$ is the photocurrent, P$_\lambda$ is the power density of incident light, A is effective device area, and $\lambda$ is excitation wavelength, reaches 10$^3$ A/W across the spectral range shown in Figure 3c. The highest value of R$_\lambda$ is 1.67 $\times$ 10$^3$ A/W at $\lambda$= 660 nm, indicating a significant photoresponse at illumination energies comparable to the bandgap of MoSSe. The high responsivity is attributed to the extended carrier lifetime resulting from the material's inherent electric field. The detectivity  
D$_\lambda$, estimated using the expression from \cite{Nayak, https://doi.org/10.1002/smll.201802593},
\begin{equation}
D_\lambda = \frac{{R_\lambda}{\sqrt{A}}}{\sqrt{2eI_{dark}}}
\end{equation}
where, I$_{dark}$ presents the dark current, reaches 2.8 $\times$ 10$^{14}$ Jones under $\lambda$ = 660 nm. This high detectivity is attributed to the low dark current of the device (Figure 3c). To asses the conversion efficiency from photon to electron, we estimated the external quantum efficiency ($\eta_{ext}$) using \cite{Nayak, https://doi.org/10.1002/smll.201802593}, 
\begin{equation}
\eta_{ext} =\frac{hcR_{\lambda}}{e\lambda}
\end{equation}
where, $h$ is Planck’s constant and $c$ is the velocity of light in vacuum, results in a maximum $\eta_{ext}$ of 3 $\times$ 10$^3$ at $\lambda$= 660 nm (Figure 3d). These results demonstrate the device's potential for broadband, high-responsivity photodetection. Moreover, the noise-equivalent power (NEP), measured at V$_{ds} = $ 2 V, reaches a minimum of 1.07 $\times$ 10$^{-18}$ W$\cdot$ Hz$^{-1/2}$ at $\lambda$=660 nm with a power density of 191 fW/$\mu$m$^2$ (Figure S9), indicating the MoSSe device sensitivity to low-light conditions. Comparative analysis of these parameters with other MoSSe alloy photodetectors (Table 1) suggests our device offers competitive advantage for practical applications. 

To understand the mode of operation of the FET device, we examined the transfer characteristics of the MoSSe phototransistor at V$_{ds} =$ 1 V under dark condition, as depicted in Figure 3e (inset). In the OFF state (V$_g <$13 V), the dark current shows only a slight increase. However, beyond a certain threshold gate voltage (V$_g >$ 13 V), the dark current increase rapidly, indicating n-type behavior due to the presence of chalcogen vacancies in the material \cite{C9MH01020C}. Upon illumination with a 633 nm wavelength (0.2 $\mu W/\mu m^2$), the device current increases across all gate voltage (Figure 3e). The calculated photo-to-dark current ratio,  $\frac{I_{photo}}{I_{dark}}$ becomes 10$^3$ at V$_{g}=$ 10 V. Additionally, the responsivity increases by 2.3 fold compared to R$_0$ (responsivity at V$_{g}=$ 0 V) at the maximum applied  gate voltage of V$_g =$ 26 V (Figure S10). These results demonstrate that gate voltage serves as an additional parameter for tuning the sensitivity of the MoSSe device.

The illumination power density plays a pivotal role in modulating the  photoresponse properties of the device. Figure 3f shows the variation of photocurrent as a function of the incident light power density ($\lambda$= 488 nm) at V$_{g}=$ 0 V. We observe that the photocurrent increases with rising power density of illumination. The dependence of photocurrent of illumination power can be expressed as $I_{illumination} \propto P^{\alpha}$.
In this case, the exponent ($\alpha$) is determined to be 0.67, which signifies that defect states are present in MoSSe \cite{Nayak, C9MH01020C}. These defect states influence photosensing behavior by trapping charge carriers, resulting in a photogating effect within the device.  Consequently, as power density increases, the values of responsivity and detectivity decrease, as  shown in Figure S11.
\begin{figure*}[t!]
\centering
\includegraphics[width=0.8\linewidth]{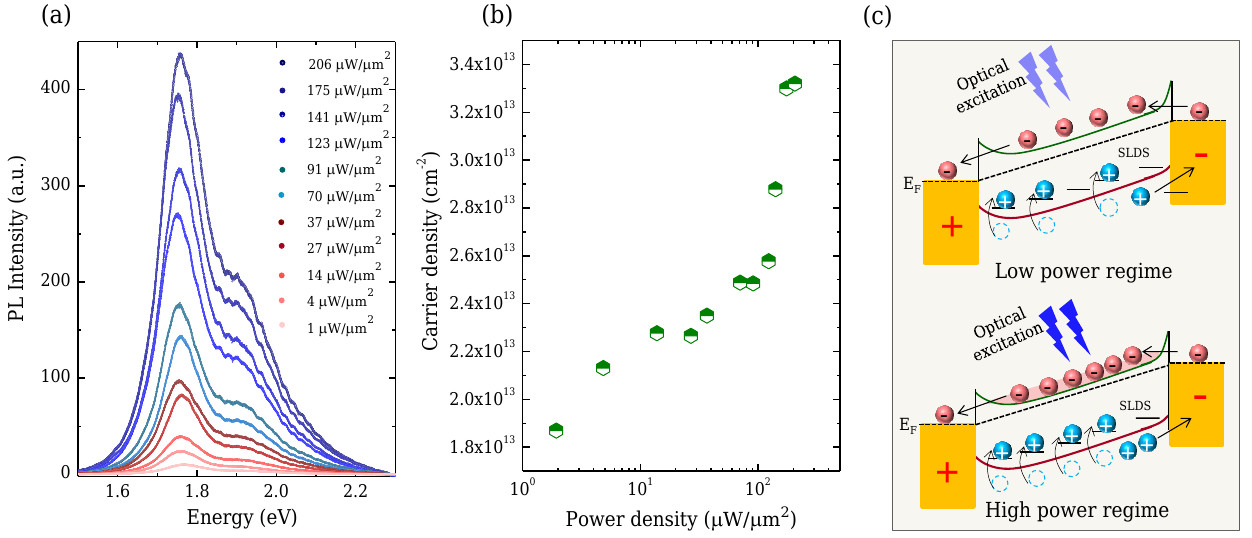}
\caption{Optical study of MoSSe. (a) PL spectra at different illumination intensities. (b) Calculated carrier density as a function of illumination intensity and (c) schematic representation of photodetection mechanism at low and high power density region.}
\label{FIG. 4.}
\end{figure*}

The temporal response of the device exhibits its capability  to follow fast-switching optical signal. The time-resolved response of the MoSSe device was measured by manually switching the light on and off every 5 seconds at V$_{ds} = $ 2 V. The response (rise and decay) curves can be modeled by using two exponential functions expressed as \cite{Masanta, PhysRevApplied.14.064029, C3TC31344A},
\begin{equation}
I_{\text{rise}}(t) = I_{\text{dark}} + A_1 \exp\left(\frac{t-t_0}{\tau_1}\right) + A_2 \exp\left(\frac{t-t_0}{\tau_2}\right)
\end{equation}

\begin{equation}
I_{\text{decay}}(t) = I_{\text{dark}} + A_1 \exp\left(-\frac{t-t_0}{\tau_1}\right) + A_2 \exp\left(-\frac{t-t_0}{\tau_2}\right)
\end{equation}
here, $A_1$, $A_2$ are scaling factors; $\tau_1$, $\tau_2$ are time constants. Optical illumination is turned on at $t = t_0$. At power density of 0.01 $\mu W/\mu m^2$, the photocurrent exhibits persistence, with a slow rise and decay time of 114 ms and 1.86 s, respectively (Figure 4a, b: bottom). However, for communication applications, photodetection requires a rapid response with minimal persistence.
\begin{table*}[t!]
\centering
\begin{tabular}{c c c c c c c}
\hline
\hline
\vspace{0.05cm}\\
 Device stucture & Method & Wavelength & Responsivity &
Detectivity & Rise/decay time & Reference\\
  &  & nm  & A/W & Jones &
 ($\tau_{rise}$/$\tau_{decay}$)   \\
 \vspace{0.05cm}\\
\hline
\vspace{0.05cm}\\
 Au/MoSSe/Au  & CVD & 532 & 2.06 & .. & 18/35 sec & \cite{https://doi.org/10.1002/adma.201901405}\\
 \vspace{0.3cm}\\
Cu/MoSSe/Cu & Hydrothermal & 660 & 0.00175 & ... & 4.7 /4.7 sec & \cite{D0NA00202J} \\
\vspace{0.3cm}\\
Au/MoSSe/Au & CVD & 660 & 1.67 $\times$ 10$^{3}$ & 2.8 $\times$ 10$^{14}$ & 18/30 ms & This work\\
\vspace{0.1cm}\\
 \hline
\end{tabular}
\caption{Parameters of  reported  alloy MoSSe based photodetectors and present work}
\end{table*}

We further investigated the temporal response of the MoSSe device under varying  illumination intensities. The rise and decay times at an illumination intensity of 159 $\mu W/\mu m^2$ are shown in Figure 4a and b (top). With increasing power density, we observe that the response time decreases for both rise and decay phases. Specifically, the rise time shortens from 114 ms to 18 ms and the decay time reduces from 1.86 s to 30 ms (Figure 4c). The resolution in Figure 4a and 4b (top panel) is limited by the measurement electronics. Beyond a certain illumination intensity, the DC measurements using a source meter were restricted because its temporal resolution required to obtain data is larger than the response time of the devices. In Figure 4c, the rise time is observed to approach a saturation value at higher illumination intensities, which can be attributed to this limitation.

Previous studies have suggested that defect states play a role in limiting the photoresponse of TMDC-based photodetectors \cite{C9MH01020C, Nur2020, PhysRevApplied.14.064029}. However, the detailed dynamics of charge trapping in these states and how illumination intensity affects temporal response remain to be fully understood. Additionally, it is still unclear if other processes contribute to photocurrent generation in TMDC-based photodetectors. In this context, we studied the PL properties of the MoSSe sample under varying illumination intensities (Figure 5a). The dependency of intensity of A, B excitons and trion with light power intensity is shown in Figure S12a, consistent with behaviors observed in other TMDC materials  \cite{Kaplan_2016, https://doi.org/10.1002/pssb.201800384}. The increasing intensity ratio of trions and A excitons (Figure S12b) with higher illumination power density suggests enhanced trion generation under illumination. Given that trions are negatively charged in our sample, as indicated by transfer characteristics, this increase in trion population implies that electron density (n$_{el}$) within the material rises with light intensity. The electron density n$_{el}$ can be calculated using the mass-action law \cite{doi:10.1021/nn504196n},
\begin{equation}
n_{el} = (\frac{4m_{A^0}m_{e}}{\pi \hbar^2 m_{A^-}}) (K_{B}T) (\frac{I_{A^-}}{I_{A^0}}) (\frac{\gamma_{A^0}}{\gamma_{A^-}}) exp(-\frac{E_b^{A^-}}{K_BT})
\end{equation} 
where, $m_{e}$ is the effective mass of the electron, $m_{A^0}= m_{e} + m_{h}$ and $m_{A^-}= 2m_{e} + m_{h}$  are the effective masses of exciton and trion respectively, $\hbar$ is the reduced Planck’s constant, $k_B$ is the Boltzmann constant, $E_b^{A^-}$is the trion binding energy, and $\gamma_{A^0}(\gamma_{A^-})$ are the radiative decay rates of exciton (trion). Figure 5b shows the estimated n$_{el}$ as a function of light intensity in the MoSSe device. To explain these findings, we propose the schematic model shown in Figure 5c. 

The n-type behavior of the MoSSe device arises from  defect states. Under illumination, photogenerated holes become trapped in these defect states, resulting in a photogating effect that enhances the photocurrent\cite{Nur2020, C9MH01020C, PhysRevApplied.14.064029}. 
At low illumination intensities, this effect is insufficient to passivate the defect states (see Figure 5c, top), leading to a slower device response. As illumination intensity increases, more holes are trapped, strengthening the photogating effect (Figure 5c, bottom). This process enhances the electron density (Figure 5b), leading to an increased trion population (Figure S12b), as evident from the PL measurements. The longer trion lifetime promotes their collection at electrodes under bias before recombination, thus boosting the photocurrent \cite{Wang2015}. Additionally, as intensity rises, more holes fill the trap states, leading to defect passivation (Figure 5c) and ultimately speeding up the device response.

 We also examined how vibrational mode changes in MoSSe using Raman spectroscopy (Figure S13a) across illumination intensities. The A$_1^2$ mode exhibits a slight red shift with increased intensity, while the E$_1^2$ peak remains largely unchanged (Figure S13b). Previous studies have shown that the A$_{1g}$ mode in MoS$_2$ is sensitive to doping, while the E$_{2g}$ mode is relatively unaffected \cite{PhysRevB.85.161403, https://doi.org/10.1002/smll.201202982}. The lack of shift in the E-type mode suggests that local heating does not significantly influence the temporal response of MoSSe. Meanwhile, the A$_1^2$ peak softening with increased illumination indicates n-type (electron) doping in the material \cite{PhysRevMaterials.5.124001}, 
 corroborating our optoelectronic observations.
\par
Our findings on the MoSSe-based photodetector, including its photoresponsivity, response time, and optical and vibrational properties under varying illumination intensities, suggest multiple mechanisms contribute to photocurrent generation. The out-of-plane electric field in MoSSe enhances the photoresponse by increasing carrier lifetime. Photogating effects, modulated by illumination intensity, switches the device from slow to fast photodetection. At higher light intensities, increased trion formation aids carrier separation due to their long lifetimes, improving charge collection and boosting photocurrent. These results demonstrate the potential of alloying to fine-tune photosensitivity in MoSSe-based devices by manipulating defect states.

\section{Conclusion}
In summary, we have demonstrated that MoSSe serves as a highly photosensitive photodetector material, operational across visible to infrared wavelengths at room temperature. The inherent out-of-plane electric field in MoSSe, confirmed by PFM measurements, polarization-resolved SHG data and theoretical first-principles calculations, enhances carrier lifetime, as supported by TRPL study, improving carrier collection at electrodes. Enhanced photosensing parameters, such as photoresponsivity, detectivity, and EQE at low power densities facilitate the detection of weak light signals with low power consumption. Additionally, the tunable temporal response with illumination intensity enables switching between slow and fast photodetection modes. Our work highlights multiple mechanisms affecting MoSSe photodetection, including the inherent electric field, photogating from hole trapping, and electron density increase with illumination, as supported by intensity-dependent PL and Raman studies. This research paves the way for developing TMDC-based optoelectronic devices with applications in miniaturized spectroscopy, spectral imaging, and threat detection.
\newpage
\section*{Supplemental Material}

\vspace{0.5 cm}
\hrule
\vspace{1 cm}
\setcounter{figure}{0}
\setcounter{equation}{0}
\renewcommand{\thetable}{S\arabic{table}}
\renewcommand{\tablename}{Table}
\renewcommand{\thefigure}{S\arabic{figure}}
\renewcommand{\figurename}{Fig.}
\renewcommand{\theequation}{S\arabic{equation}}

\subsection*{Section I. Methods}
\subsection*{Chemical vapor deposition (CVD) synthesis of MoSSe sample}
We have chosen CVD method to synthesis alloy MoSSe samples on SiO$_2$/doped Si substrates in a homemade horizontal tube furnace connected to a vacuum pump, as illustrated in Figure S1.  Molybdenum trioxide (MoO$_3$), sulfur, and selenium powders were taken as precousers for this synthesis. Molybdenum trioxide (MoO$_3$, 267856-100G) and sulfur (414980-50G) were obtained from Sigma-Aldrich, while selenium powder (36208) was sourced from Alfa Aesar. The substrates were cleaned using a piranha solution and water. To grow MoSSe alloy on substrates, we optimised the synthesis parametrs including distances between the Mo, S, and Se sources, temperature, and flow rates of Ar gas. The MoO$_3$ powder, placed in a quartz boat, was positioned in the high-temperature zone (approximately 750$^\circ$C), while S and Se powders were placed in a lower temperature zone (around 200$^\circ$C). The furnace temperature was increased at a rate of 25$^\circ$C/min under a flow of 10 sccm of high-purity Ar gas, reaching the maximum temperature over 30 minutes. The furnace was held at this temperature for 10 minutes under a 40 sccm Ar flow, followed by natural cooling to room temperature. 

\begin{figure*}[h!]
\centering
\includegraphics[width=1\linewidth]{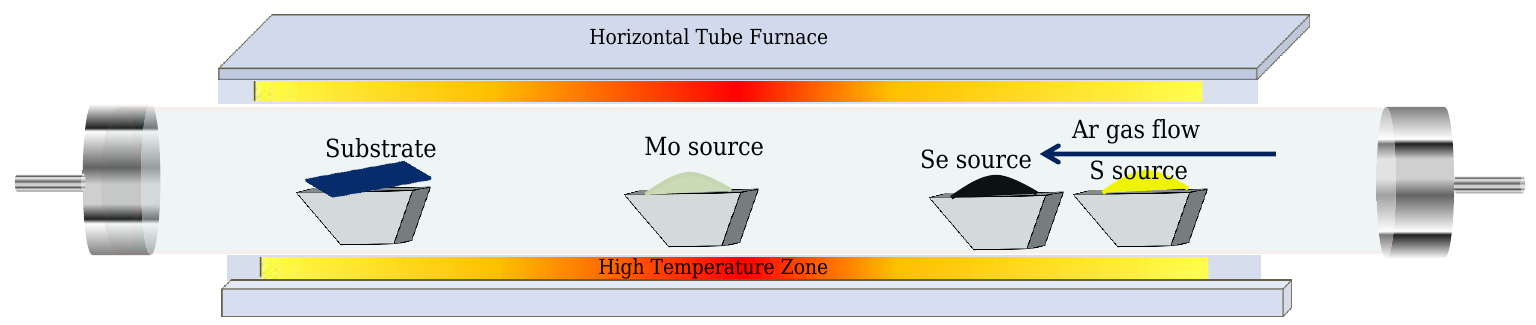}
\caption{CVD synthesis of alloy MoSSe.}
\label{FIG. 4.}
\end{figure*}
\subsection*{Primary characterizations}
Atomic force microscopy (AFM) was conducted using a Bruker Innova instrument to determine the layer number of the as-synthesized samples. Compositional analysis was performed with an energy dispersive X-ray (EDX) spectrometer integrated with an FEI Quanta 200 scanning electron microscope. We performed second harmonic generation (SHG) measurements on monolayer MoSSe to study nonlinear optical response, as discussed in details in Section III. Piezoelectric force microscopy (PFM) was performed in contact mode using an AFM equipped with a dual AC resonance tracking piezoresponse module (MFP-3D™, Asylum Research). The cantilevers used in these measurements had spring constants of approximately 2.8 N/m and free-air resonances at 75 kHz and the used Si probes were coated with Ti/Ir to ensure electrical conductivity. In order to study the optical and vibrational properties of MoSSe sample, PL and Raman measurements were performed using a micro-Raman spectrometer (LabRAM HR, Horiba JobinYvon) equipped with a Peltier-cooled CCD detector. The experiments were done using an air cooled argon-ion laser (Ar+) with a wavelength of 488 nm as an excitation light source. Time-resolved photoluminescence (TRPL) spectroscopy was carried out using a 375 nm diode laser (NanoLED, HORIBA Scientific) with a pulse width of 50 ps and a repetition rate of 500 kHz as the excitation source. The emitted PL signal was dispersed by a double-grating monochromator and detected using a picosecond photon counter (HORIBA Jobin-Yvon). All the measurements were performed at room temperature. 

\subsection*{Computational details for first-principles calculations}

First-principles calculations were performed within the framework of density functional theory as implemented in the Quantum ESPRESSO package\cite{Giannozzi_2009}. The energy and charge density cutoff values for the expansion of the plane-wave basis set were chosen to be 50 Ry and 500 Ry, respectively.
Electron-electron interactions were treated using the generalized gradient approximation (GGA) as the exchange-correlation functional\cite{GGA}. Electron-ion interactions and spin-orbit coupling were treated with fully relativistic effects using projector-augmented wave pseudopotentials\cite{PAW}. The Brillouin-zone sampling was performed using a $\Gamma$-centered Monkhorst-Pack k-point grid of 24$\times 24 \times 1$. A 2$\times 2\times 1$ monolayer polar TMDC cell models the MoSSe alloy  with 38\% Se and 62\% S. The optimized in-plane lattice parameter is 6.42 \AA. Atomic positions are relaxed until forces fall below 0.1 mRy/Bohr.

\subsection*{Device fabrication and optoelectronic measurements}
To prepare electrical contacts, the as-grown monolayer MoSSe samples are cleaned using standard technique (acetone and IPA cleaning). Next, the samples were spin-coated with AZ1512-HS positive photoresist and baked at 100$^\circ$C for one minute and a laser writer (Microtech, Model LW405) with a laser power of 160 mJ/cm$^2$ was utilized to design the pattern for the electrical contacts. Following the patterning step, a deposition of Cr/Au (5 nm/60 nm) was carried out using an e-beam evaporation system. The channel length and width were kept 5 $\mu m$ and 6 $\mu m$, respectively. The entire device fabrication process are conducted meticulously to prevent any contamination. To examine photoresponse feature of the device, we have used Bentham spectral response measurement setup equipped with 75 W Xenon/QTH lamp as light source, Bentham TMC 300 monochromator and a lock-in amplifier (496 DSP). The DC measurements were performed using keithley 2450 source meter. Electrical measurements were performed at room temperature.

\subsection*{Section II. EDX and Raman measurements of MoSSe}

\begin{figure*}[h!]
\centering
\includegraphics[width=0.8
\linewidth]{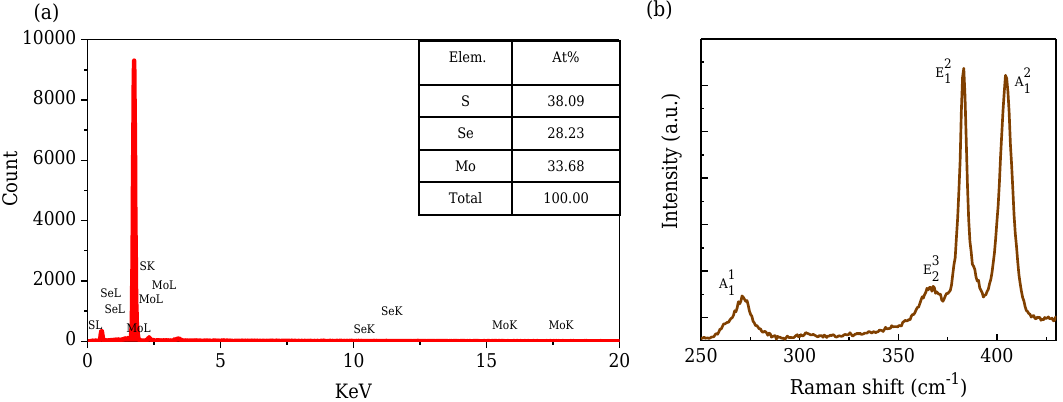}
\caption{(a) EDX spectrum of the monolayer MoSSe. Inset table provides the estimated percentages of Mo, S, and Se from this study and (b) Raman spectrum of monolayer MoSSe.}
\label{FIG. 4.}
\end{figure*}

\subsection*{Section III. SHG Measurement}
The experimental setup for polarization-resolved SHG measurements in a reflection geometry is shown in Figure S3. A Ti:sapphire laser, emitting 100 fs pulses at a repetition rate of 80 MHz, with a central wavelength of 800 nm, serves as the light source for the experiment. The input laser beam is initially passed through a neutral density (ND) filter for adjusting the incident laser power.
\begin{figure*}[h!]
\centering
\includegraphics[width=0.8
\linewidth]{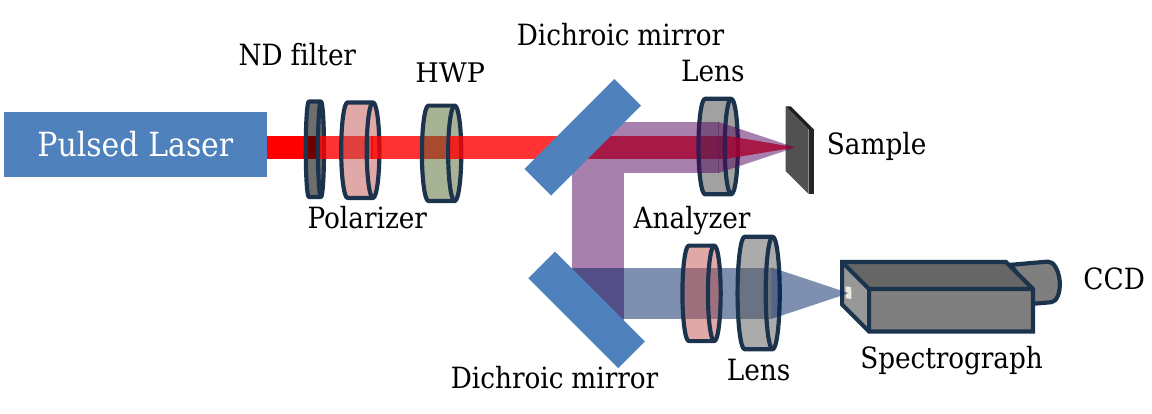}
\caption{Schematic representation of polarization-resolved SHG experimental setup}
\label{FIG. 4.}
\end{figure*}
After the ND filter, the beam passes through a linear polarizer, followed by a half-wave plate (HWP) to rotate the polarization direction of the incident light. The input near-infrared (NIR) beam passes through a dichroic mirror, which reflects the blue SHG signal while transmitting the NIR input beam. The input laser beam is then focused onto the sample using a plano-convex lens with a focal length of 6 cm.

The SHG signal generated by the sample is captured by the same lens and reflected by the dichroic mirror. The signal then passes through a second dichroic mirror, which further filters the signal, ensuring that only the SHG component is transmitted. The filtered SHG signal is focused onto a monochromator via a lens with a focal length of 7 cm, with an analyzer (co/cross configuration to the pump excitation) placed in the path to resolve polarization information. The SHG signal, after dispersion in the monochromator, is detected by a thermoelectrically cooled CCD camera. The average pump power density used in the experiment was 0.7 mW/$\mu$m$^2$. 

\begin{figure*}[h!]
\centering
\includegraphics[width=0.7
\linewidth]{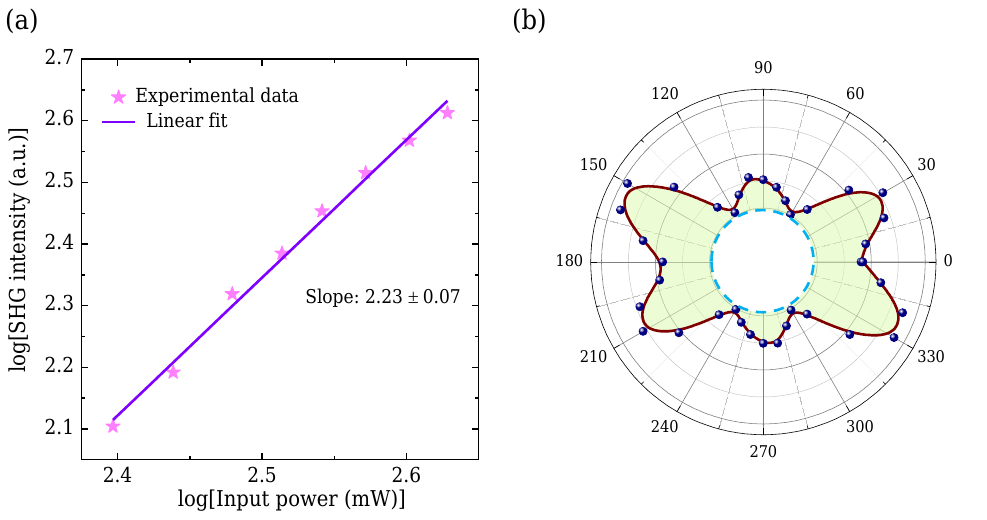}
\caption{(a) SHG intensity as a function of pump power, with experimental data represented by pink stars and the fitted curve shown as a solid purple line and
(b) polar plot of the polarization-resolved SHG intensity, measured with the detection polarizer maintained perpendicular to the pump polarization.}
\label{FIG. 4.}
\end{figure*}
\par
We performed power dependent SHG study (Figure S4a) which confirms the second-order non-linear properties arising from our as-synthesized sample. Additionally, polarization-resolved SHG pattern (Figure S4b) exhibits a six-lobed pattern, obtained by rotating the pump polarization using a half-wave plate, while maintaining the detection polarizer perpendicular to the pump. This pattern indicates that the in-plane emission dipole ($\chi_{yyy}$) contributes to the six-lobed pattern, whereas the out-of-plane emission dipole ($\chi_{zxx}$) accounts for the nonvanishing, isotropic background SHG intensity, as marked by blue dotted circle in Figure S4c \cite{doi:10.1021/acsnano.4c02854, https://doi.org/10.1002/adom.202300958}. 

\newpage
\subsection*{Section IV. PFM analysis monolayer MoSSe and substrate}
\begin{figure*}[h!]
\centering
\includegraphics[width=0.6
\linewidth]{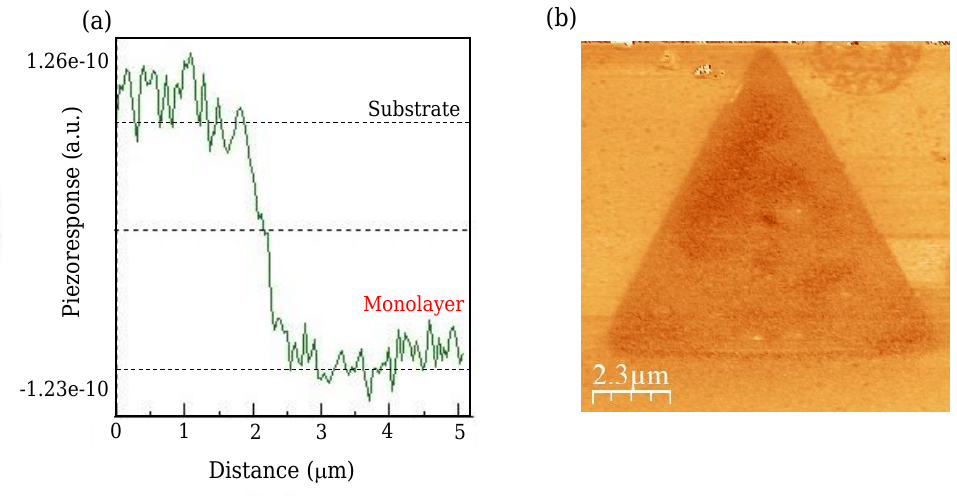}
\caption{(a) PFM amplitude plot and (b) phase image of monolayer MoSSe}
\label{FIG. 4.}
\end{figure*}

\begin{figure*}[h!]
\centering
\includegraphics[width=0.6
\linewidth]{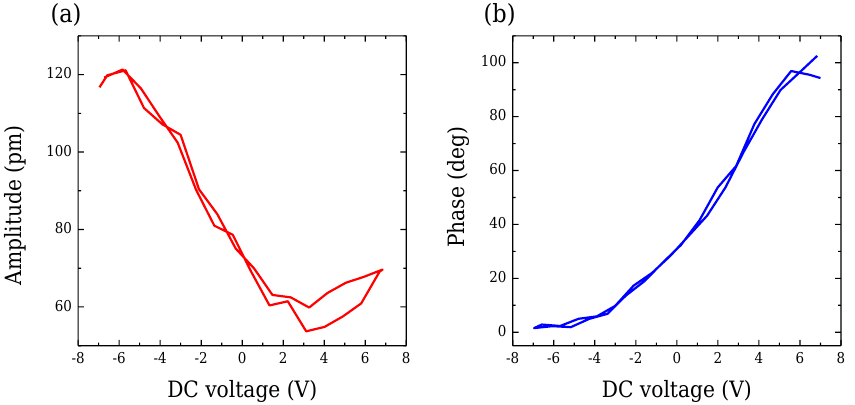}
\caption{PFM measurement on Si/SiO$_2$ substrate. (a) PFM amplitude and (b) phase as a function of DC voltage.}
\label{FIG. 4.}
\end{figure*}

\newpage
\subsection*{Section V. DFT calculated electrostatic potential and electronic band structure of monolayer MoS$_2$ ad MoSSe}
The electrostatic potential with respect to the vacuum level is plotted in Figure S7 along the direction perpendicular to the plane of MoS$_{2}$ monolayers.
Figure~\ref{FIG.4} shows the fully-relativistic electronic band structure of (a) MoS$_{2}$ and (b) MoSSe with spin-orbit coupling. Rashba splitting around the high-symmetry point $\Gamma$ is observed for MoSSe due to the presence of intrinsic electric field, which is absent for MoS$_{2}$, see the insets.
\begin{figure*}[h!]
\centering
\includegraphics[width=0.35
\linewidth]{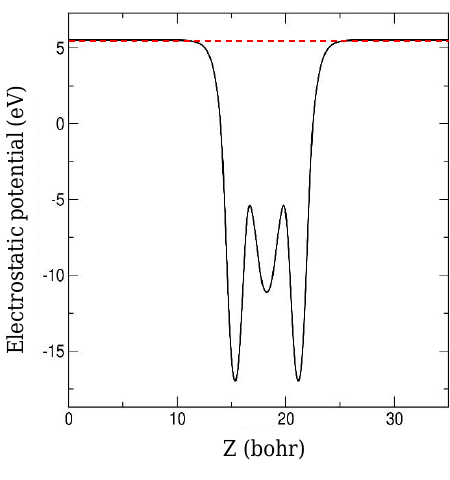}
\caption{Electrostatic potential with respect to the vacuum level is plotted along the direction perpendicular to the plane of MoS$_{2}$ monolayers}
\label{FIG.4}
\end{figure*}

\begin{figure*}[h!]
\centering
\includegraphics[width=0.6
\linewidth]{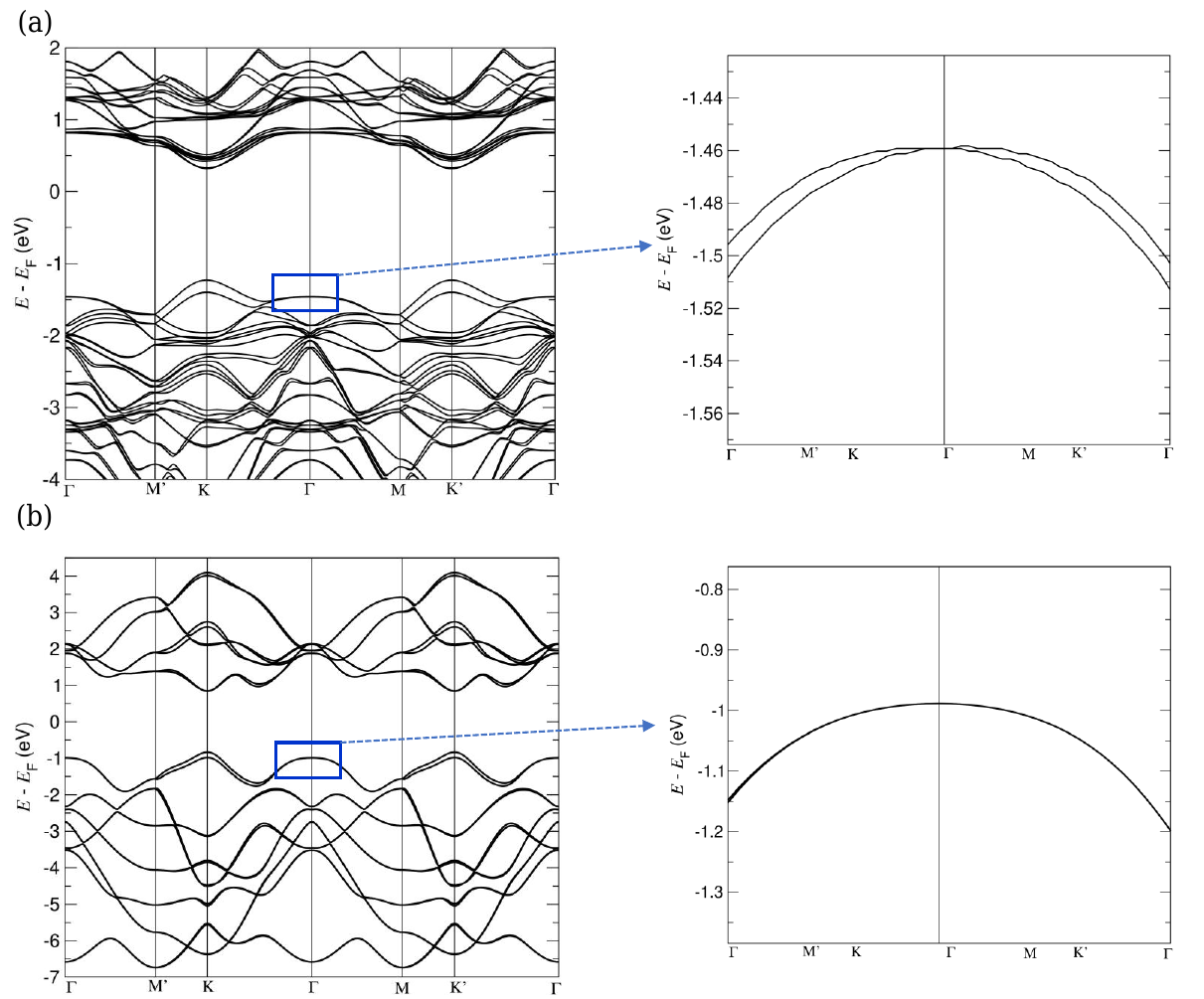}
\caption{(a) and (b) DFT-calculated electronic band structures of monolayer MoSSe and MoS$_2$, respectively. The insets of (a) and (b) show the magnified regions near the $\Gamma$ point. In the case of MoS$_2$, no Rashba splitting is observed, whereas for MoSSe, a clear Rashba splitting appears near the $\Gamma$ point.}
\label{FIG.4}
\end{figure*}

\newpage

\subsection*{Section VI. Calculation of noise-equivalent power (NEP)}
We computed the noise-equivalent power (NEP) by solely considering the impact of shot noise, utilizing the subsequent formula \cite{Nayak,doi:10.1021/acsphotonics.8b00318},
\begin{equation}
 NEP =\frac{\sqrt{2eI_{dark}}}{R_\lambda}
\end{equation}
The estimated values of NEP at V$_{ds} = $ 2 V is 1.07 $\times 10^{-18}$ W$\cdot$ Hz$^{-1/2}$ at $\lambda$=660 nm with power density 191 fW/$\mu$m$^2$, as depicted in Figure S3.
\begin{figure*}[h!]
\centering
\includegraphics[width=0.4\linewidth]{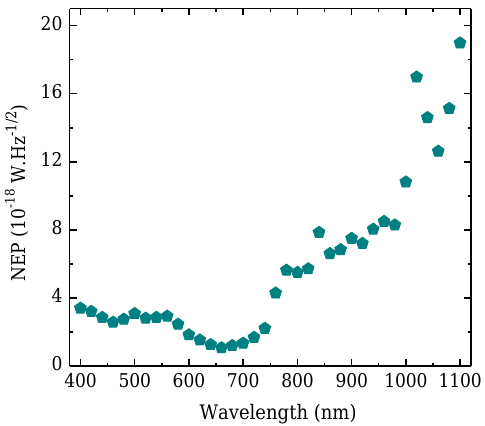}
\caption{Calculated NEP at V$_{ds} =$ 2 V over the broad spectral region (400-1100 nm).}
\label{FIG. 4.}
\end{figure*}
\newpage
\subsection*{Section VII. Gate voltage dependent responsivity plot}
\begin{figure*}[h!]
\centering
\includegraphics[width=0.4\linewidth]{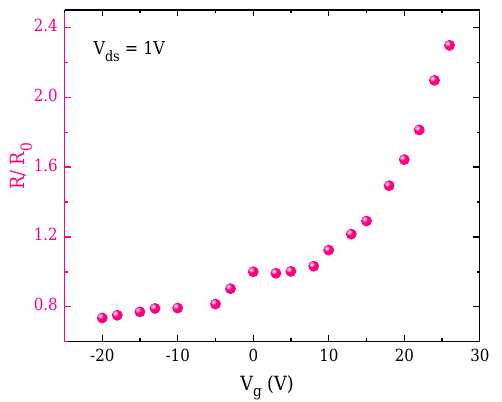}
\caption{Plot of $R/R_0$ ( where $R$ and $R_0$ are responsivities at any V$_g$ (- 20 V$\leq$V$_g$ $\leq$ + 26 V) and V$_g$= 0 V respectively) as a function of gate voltage.}
\label{FIG. 4.}
\end{figure*}


\subsection*{Section VIII. Power dependent responsivity and detectivity plot}
\begin{figure*}[h!]
\centering
\includegraphics[width=0.4\linewidth]{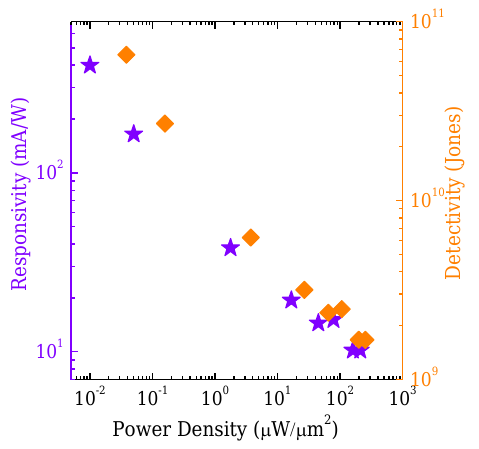}
\caption{Logarithmic representation of responsitivity and detectivity as a function of power density.}
\label{FIG. 4.}
\end{figure*}
\newpage
\subsection*{Section IX. Extracted parameters from illumination intensity dependent PL study}
\begin{figure*}[h]
\centering
\includegraphics[width=0.7\linewidth]{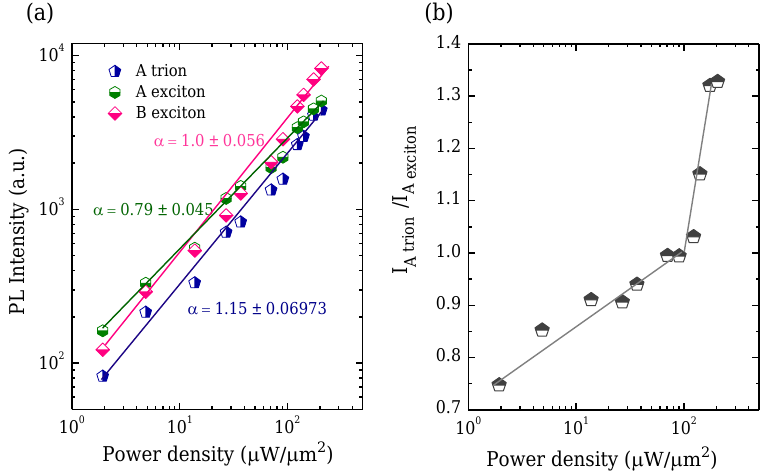}
\caption{Extracted parameters from PL study of MoSSe. (a) logarithmic representation of intensity of PL peaks and their power law fitting. (b) Ratio of trion to A exciton as a function of illumination intensity.}
\label{FIG. 4.}
\end{figure*}

\subsection*{Section X. Illumination intensity dependent Raman study}
\begin{figure*}[h!]
\centering
\includegraphics[width=0.6\linewidth]{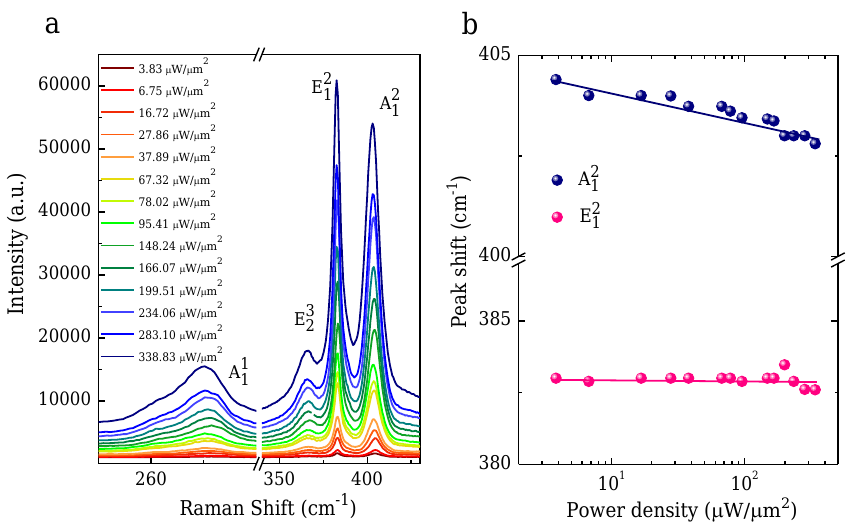}
\caption{Raman study of MoSSe. (a) Raman spectra at different illumination intensities and (b) peak shift of Raman modes as a function of illumination intensity.}
\label{FIG. 4.}
\end{figure*}

\newpage
%

\subsection*{Acknowledgements} 
We sincerely acknowledge Prof. Indrani Bose, Bose Institute, for her enriching discussions and insightful feedback.
C. Nayak acknowledges the INSPIRE Fellowship Programme, DST, Government of India, for granting her a research fellowship with Registration Number IF180057. S. Masanta expresses gratitude to the Council of Scientific $\&$ Industrial Research (CSIR), India, for the financial support provided through the NET-SRF award (File Number 09/015(0531)/2018-EMR-I). S. Ghosh acknowledges the computational resources provided by the Swedish National Infrastructure for Computing (SNIC) at NSC, PDC, and HPC2N, which were partially funded by the Swedish Research Council. S. Ghosh also acknowledges the supercomputing facility received from the National Supercomputing Mission (NSM) through PARAM Seva located at IIT Hyderabad implemented by C-DAC and supported by the Ministry of Electronics and Information Technology (MeitY) and the Department of Science and Technology (DST), Government of India. B. Pal acknowledges financial support from the SERB, India (File Number CRG/2022/008885). A. Singha acknowledges financial support from the Science and Engineering Research Board (SERB), India (File Number EMR/2017/002107).

\end{document}